\documentclass[aps,prd,twocolumn,superscriptaddress,groupedaddress,nofootinbib]{revtex4-1} 
\usepackage{amsmath,amssymb}
\usepackage{epsfig}
\usepackage{comment}
\usepackage{graphicx}
\usepackage[usenames,dvipsnames]{color}
\usepackage{slashed}
\usepackage{amsmath}
\usepackage{color}

\usepackage{tikz-feynman}
\tikzfeynmanset{compat=1.1.0}

\usepackage{tikzsymbols}

\begin{document}
\title{Leptogenesis in  supersymmetry  with one $L$ violating coupling }
\author{Rathin Adhikari}
\email{rathin@ctp-jamia.res.in}
\affiliation{Centre for Theoretical Physics, Jamia Millia Islamia (Central University), \\ New Delhi 110025, India}
\author{Arnab Dasgupta}
\email{arnabdasgupta@protonmail.ch}
\affiliation{School of Liberal Arts, Seoul-Tech, Seoul 139-743, Korea}

\begin{abstract}
We have shown a new scenario of successful leptogenesis with one $L$ violating coupling and a relative Majorana phase playing the role of $CP$ violation. This is in contrast to the usual consideration of Feynman diagram with at least two $L$ violating couplings.  We have considered $R$-parity violating Minimal Supersymmetric Standard Model (MSSM) for leptogenesis at TeV scale. This scenario is also consistent with generating  light neutrino mass if asymmetry is generated through semileptonic $\lambda^{\prime}$ coupling. 

\hspace*{\fill} 


\end{abstract}

\maketitle

\section{Introduction} There is asymmetry in the number density of baryons ($n_B$ ) and antibaryons ($n_{\bar{B}}$) in our observed universe \cite{asym} and  
one may consider baryogenesis or leptogenesis mechanism \cite{early}  to explain such asymmetry. In the latter case the lepton number asymmetry could result in baryon number asymmetry in presence of sphalerons. 

For successful baryogenesis/leptogenesis mechanism three basic
Sakharov's conditions   \cite{sakh}:  (1) Presence of baryon number ($B$) or lepton number ($L$) violating interactions (2) $C$ and $CP$ violating physical process and (3) departure of those physical processes from thermal equilibrium, 
are to be fulfilled. Non-zero $CP$ asymmetry requires interference of tree level and higher order Feynman diagrams related to those physical process. However, to get non-zero $B$ or $L$ asymmetry certain conditions \cite{nano,raghu} on  higher order Feynman diagram are to be satisfied. 
According to
Nanopoulos-Weinberg theorem \cite{nano}  such  diagram should require at least two $B$ or $L$ violating couplings for the decaying particle having only $B$ or $L$ violating decay modes \cite{Sorbello:2013xwa}. As considering more $B$ or $L$ violating couplings in Feynman diagram reduces the asymmetry due to smallness of such couplings, it would be more appropriate to find out  baryogenesis or leptogenesis mechanism  where restriction of this theorem could be avoided. This could be possible if the decaying  particle generating asymmetry has $B$ or $L$ conserving decay modes also \cite{raghu}. We have considered leptogenesis mechanism in the context of $R$ parity violating Minimal Supersymmetric Standard model where next to lightest neutralino could have this property of having both $L$ violating and $L$ conserving decay modes as shown in figure \ref{fig1} and figure \ref{fig2}. Here,
we have considered  only one $L$ violating coupling in figure \ref{fig1} to produce $L$ asymmetry
which does not contradict the Nanopoulos-Weinberg theorem \cite{raghu}. In case of asymmetry 
generated by lightest neutralino which does not have  $L$ conserving decay modes, there should be
more than one $L$ violating coupling. As for example, in one work \cite{Masiero:1992bv} related to lightest neutralino decay three $L$ violating couplings are required  in higher order diagram.

\begin{figure}[th]
\centering
\begin{tabular}{cc}
\begin{tikzpicture}[/tikzfeynman/small]
\begin{feynman}
\vertex (i){$\chi^0_1$};
\vertex [right = 1.2cm of i] (j);
\vertex [above right =1cm of j] (a){$l,u$};
\vertex [right =1.4cm of j] (e) ;
\vertex [above right =1cm of e] (f) {$u,l$};
\vertex [right =1cm of e] (g) {$\bar{d}$};
\diagram*[small]{(i) -- [anti fermion] (j),(j) -- [fermion] (a),(g) -- [fermion] (e) -- [charged scalar,edge label={$\tilde{l},\tilde{u}$}] (j),(e) --[fermion] (f)};
\end{feynman}
\end{tikzpicture} &
\begin{tikzpicture}[/tikzfeynman/small]
\begin{feynman}
\vertex (i){$\chi^0_1$};
\vertex [right = 1.2cm of i] (j);
\vertex [above right =1cm of j] (a){$\bar{d}$};
\vertex [right =1.4cm of j] (e) ;
\vertex [above right =1cm of e] (f) {$l$};
\vertex [right =1cm of e] (g) {$u$};
\diagram*[small]{(i) -- [fermion] (j),(j) -- [anti fermion] (a),(g) -- [anti fermion] (e) -- [anti charged scalar,edge label=$\tilde{d}$] (j),(e) --[fermion] (f)};
\end{feynman}
\end{tikzpicture}
\end{tabular}
\begin{tikzpicture}[/tikzfeynman/small]
\begin{feynman}
\vertex (i){$\chi^0_1$};
\vertex [right = 1.2cm of i] (j);
\vertex [above =0.9cm of j] (c1);
\vertex [right =1.1cm of c1] (a1);
\vertex [below =1.4cm of a1] (b1) ;
\vertex [above right =1cm of j] (a);
\vertex [above right =1cm of a] (b){$l,u$};
\vertex [below right =1 cm of a] (d);
\vertex [right =1.4cm of d] (e) ;
\vertex [above right =1cm of e] (f) {$u,l$};
\vertex [right =1cm of e] (g) {$\bar{d}$};
\diagram*[small]{
(i) -- [fermion] (j),(j) -- [charged scalar,edge label={$\tilde{l},\tilde{u}$}] (a),(a) -- [fermion] (b), (a) -- [majorana,edge label=$\chi^0_2$] (d),(j) -- [anti fermion,edge label' = {$l,u$}] (d),(g) -- [fermion] (e) -- [charged scalar,edge label={$\tilde{l},\tilde{u}$}] (d),(e) --[fermion] (f),(a1) -- [scalar,style={thick,red}] (b1)};
\end{feynman}
\end{tikzpicture}
\caption{The tree level and next higher order diagrams  giving rise to 
leptonic asymmetry. Lepton number violating coupling is on the right of the 
vertical `cut' shown in second diagram\footnote{Readers may note that 
contrary to standard notations, we consider
$\chi^0_1$ as next to lightest neutralino and 
$\chi^0_2$ as the lightest neutralino.}
}

\label{fig1}
\end{figure}

\begin{figure}[th] 
\centering
\begin{tabular}{c}
\begin{tikzpicture}[/tikzfeynman/small]
\begin{feynman}
\vertex (i){$\chi^0_1$};
\vertex [right = 1.2cm of i] (j);
\vertex [above right =1cm of j] (a){$l,u,d$};
\vertex [right =1.4cm of j] (e) ;
\vertex [above right =1cm of e] (f) {$\bar{l},\bar{u},\bar{d}$};
\vertex [right =1cm of e] (g) {$\chi^0_2$};
\diagram*[small]{(i) -- [anti fermion] (j),(j) -- [fermion] (a),(g) -- [fermion] (e) -- [charged scalar,edge label={$\tilde{l},\tilde{u},\tilde{d}$}] (j),(e) --[anti fermion] (f)};
\end{feynman}
\end{tikzpicture}
\\
\end{tabular}

\caption{$L$ conserving decays of $\chi^0_1$.}
\label{fig2}
\end{figure}

 Another condition on higher order diagram is as follows:  The interaction between intermediate on-shell particles and the final particles  should correspond to a net change in baryon/lepton number   \cite{raghu}. In other words, $B$ or $L$ violating coupling should be present on the right of the `cut' in the higher order diagram  (`cut' on the internal line is possible when on-shell condition is satisfied indicating presence of imaginary part of loop integral which is  required for $CP$ asymmetry) with net change in $B$ or $L$ due to all such couplings on the right with final states. This      is essential to have non-vanishing total  $B$ or $L$ asymmetry after summing over all possible intermediate and final states generated due to decay.  One may get non-zero asymmetry from the interference of tree and higher order diagram without right of the `cut' condition, but it would vanish if  other diagrams with all possible intermediate and final states are considered. The cases where 
 some authors have found vanishing of total  asymmetry \cite{fonggarbrecht} are 
 examples of this feature.  In our case in leptogenesis,  above-mentioned condition is satisfied in the higher order Feynman diagram 
 for neutralino decay as shown in figure \ref{fig1}.

\section{Low Scale Leptogenesis in MSSM} 
We have shown  in  $R$ parity violating MSSM, leptogenesis could be possible   at low energy scale ($\mathcal{O}$ TeV). There are some interesting works on low scale  leptogenesis \cite{chun} considering non-supersymmetric models \cite{nonsusy,smi,gar} and supersymmetric models
\cite{Masiero:1992bv,Sarkar:1996sn,Hambye:2000zs,Hambye:2001eu,gross,sheng1,Cui:2013bta}. In non-supersymmetric models, in resonant leptogenesis scenario \cite{nonsusy} the asymmetry is enhanced by resonance through self-energy effects. But certain amount of fine tuning is required.  At least, masses of the two heavy neutrinos are required to be quasi-degenerate with their mass difference of the order of their decay widths. In heavy neutrino oscillation scenario \cite{smi} of leptogenesis, the Yukawa couplings of electroweak singlet neutrinos are required to be very small ($\sim 10^{-8}$) and near mass degeneracy of heavy neutrinos are required. However, it was shown later \cite{gar} that such mass degeneracy may not be required but then the Yukawa couplings are somewhat higher than what is
in general, considered for most of the points in parameter space for see-saw mechanism.  In  works related to supersymmetric models,   the minimal version of the model - the MSSM with explicit $R$ parity violation \cite{suz}, seems to be inadequate for the generation of asymmetry. Due to that it  has been modified  in the scalar sector by introducing new scalar field or has been modified by considering the $R$ violating interactions with non-holomorphic terms in the superpotential. Even for leptogenesis at high scale ($\sim 10^{6}$ GeV), in  interesting soft leptogenesis scenario
in non-minimal supersymmetric models, some fine-tuning is required with masses of sneutrinos to
avoid the gravitino over-abundance problem \cite{gross}. Otherwise,  generic trilinear soft supersymmetry breaking couplings are to be assumed \cite{sheng1}. However, here we have shown the possibility of successful leptogenesis at low energy scale in
the minimal version itself,  which is devoid of aforementioned problems.

In MSSM with explicit $R$-parity violation \cite{suz}, $L$ violation could come from the following trilinear R-parity violating terms in the superpotential:
\begin{align}
W_{RPV} &= \sum_{i,j,k}\left(\frac{1}{2}\lambda_{ijk}L_iL_jE^c_k + \lambda^{\prime}_{ijk}L_i Q_jD^c_k\right)   \;  , 
\end{align} 
where $L_i$ and $Q_i$ are $SU(2)$ doublet lepton and quark superfield respectively and $E_i$ and $D_i$ are 
$SU(2)$ singlet charged lepton and down type quark superfield respectively.  We have assumed that the bilinear $L$ violating term in the superpotential could be rotated away with suitable field redefinition. However, in general, such couplings are expected to be very small due to the cosmological bound on the light neutrino masses  \cite{cons}.   We have considered only the presence of non-zero $\lambda^{\prime}_{ijk}$ couplings in our numerical analysis. Similar results are expected to follow if one considers both $\lambda_{ijk}$ and $\lambda^{\prime}_{ijk}$ couplings  or any one of those couplings as non-zero for the generation of leptonic and hence baryonic asymmetry. However, if one considers any of these or both such couplings for the generation of active neutrino mass then those neutrinos are of Majorana type. In that case,  lepton numbers of neutrinos are not well-defined. But in generating leptonic asymmetry with $\lambda_{ijk}$ couplings through the decays of neutralino, lepton number of neutrinos and antineutrinos are required. But for $\lambda^{\prime}_{ijk}$ couplings,  such role is played by the lepton number of charged leptons and anti-leptons. So if same coupling is assumed to be connected with leptonic asymmetry as well as neutrino mass then  that should be $\lambda^{\prime}_{ijk}$ couplings as considered in this work.

In our scenario the leptonic asymmetry is generated above electroweak symmetry breaking scale. In that case,  the next to lightest neutralino (which is only bino without any wino or Higgsino component) has only three body decay modes. The decay of next to lightest neutralino ($\chi^0_1$) occurs through $L$ violating decay mode $\chi^0_1
\rightarrow u l \overline{d}$ and $L$ conserving decay mode $\chi^0_1
\rightarrow \chi^0_2 l \overline{l}(q\overline{q})$ . The decay process $\chi^0_1
\rightarrow u l \overline{d}$ with $\Delta L = 1$  and its conjugate process  $\chi^0_1
\rightarrow \overline{u} d \overline{l}$  generates asymmetry with $\lambda^{\prime}_{ijk}$ coupling  . For  $L$ violating decay the tree level diagrams and next higher order Feynman diagrams  with one $L$ violating couplings are shown  in figure 1 in which  charged left slepton $\widetilde{l}$ and $\widetilde{u}$ are the superpartner of left charged lepton $l$ and left $u$ quark respectively and $\widetilde{d}$ is the superpartner of right $d$ quark. In the loop diagram, there is no right $\widetilde{d}$ in the internal line, as this does not interact with $\chi^0_2$. The $L$ conserving decays are shown in figure 2  which are mediated by left as well as right squarks and  sleptons. There is also diagram with $l, u, d$ and $\bar{l}, \bar{u}, \bar{d}$ interchanged with opposite arrows which is not shown.  

We consider the neutralino mass matrix above electroweak symmetry breaking scale where the leptonic asymmetry is expected to be converted to baryonic asymmetry in presence of sphalerons as discussed later. The mass matrix is given by 
\begin{align}
\mathcal{M}_{\chi^0} &= \begin{pmatrix}
M_1 & 0 & 0 & 0 \\ 
0 & M_2 & 0 & 0 \\
0 & 0 & 0 & -\mu \\
0 & 0 & -\mu & 0
\end{pmatrix} &; \quad \begin{matrix} 
m_{\chi^0_1} = |M_1| \\ 
m_{\chi^0_2} = |M_2| \\
m_{\chi^0_3} = |\mu| \\
m_{\chi^0_4} = |\mu| \\
\end{matrix}
\end{align}
where $\mu$ is the Higgsino mass parameter and $M_1$ and $M_2$ are the $U(1)$ and $SU(2)$ gaugino mass parameters respectively. One may note that there are no gaugino-Higgsino mixing in the above mass matrix as that occurs due to electroweak symmetry breaking when  Higgs scalar  is replaced by its vacuum expectation value  in Higgs-gaugino-Higgsino interaction. We have  considered $|\mu| > (|M_1|,|M_2|) $ as otherwise  next to lightest neutralino would decay dominantly to  higgs and lightest neutralino. Then it will be hard to get enough leptonic asymmetry from its $L$ violating decay modes.   Also we consider  $|M_1| > |M_2|$ in our work as otherwise for  $|M_1| < |M_2|$ there would be  annihilation of the next to lightest neutralino   to $W^+ W^-$. 
 The dominant decay channel for $\chi^0_{3,4}$ corresponds to decaying to lightest higgs and other lighter neutralinos and as such the leptonic asymmetry generated by these are expected to be negligible. Furthermore, any asymmetry generated by these or generated at high scale by the $L$ violating decay of heavy particles will be washed out by the asymmetry generated at low scale by the decay of  lighter neutralino. The lighter two neutralinos do not decay to Higgs scalar because of the above structure of neutralino mass matrix. $\chi^0_{2}$ being the lightest neutralino for above mass matrix, has no $L$ conserving decay mode and  has only $L$  violating decay modes. So according to our earlier discussion, the next to lightest neutralino $\chi^0_{1}$ with both $L$ conserving and $L$ violating decay modes, turns out to be the only suitable neutralino for creating sufficient asymmetry with one
 $L$ violating coupling.

To obtain asymmetry, $CP$ violation is also required. In our work, this comes due to Majorana nature of neutralino fields. For $CP$ violating phase, the relevant part in the above complex symmetric neutralino mass matrix, is the upper left $2 \times 2$ block. This block contains complex mass parameters $M_1$ and $M_2$ . This block is decoupled from the rest part of mass matrix so far diagonalisation is concerned. So, lighter two neutalinos $\chi_1$ and 
$\chi_2$ states could contain  at  least one relative Majorana phase $e^{-i\phi}$ as diagonal phase matrix in the neutralino mixing matrix. This is unlike CKM mixing matrix in quark sector where three fields are necessary for $CP$ violating phase. This Majorana phase will give non-zero $CP$ violating phase  $e^{-i 2\phi}$ in the amplitude of second diagram shown in Figure \ref{fig1}. This happens due to the presence of clashing arrows on propagator $\chi_2$ consistent with its Majorana nature.  Now the asymmetry comes from the interference of tree and next higher order diagram as shown in Figure 1 and this 
phase gives a factor of $\sin 2 \phi$ in the $CP$ asymmetry. One could have considered such Majorana phases
to be present in $L$ conserving MSSM couplings also.
$L$ violating complex $\lambda^{\prime}$ coupling do not contribute to any further $CP$ violation in generating asymmetry from the decay of $\chi_1$.   
In the triangle diagram in figure \ref{fig1}, $\chi_1^0$ is not possible in the internal line as in that case, there will be no $CP$ violating phase in the interference term.  The other two heavier neutralinos in the internal line in the triangle diagram
 could not be considered as the `cut' in the diagram required for asymmetry will not be possible. So $\chi_2^0$ on the internal line 
 in the higher order diagram is the only option in our case. The decay of lightest neutralino $ \chi^0_2$ will not generate asymmetry through such diagrams with one $L$ violating coupling as $CP$ violating phase in the interference term as well as `cut' in the diagram will not be possible  simultaneously. This is consistent with our earlier discussion related to Nanopoulos-Weinberg theorem as $ \chi^0_2$ has no $L$ conserving decay mode. However, the decay of $ \chi^0_2$ will contribute to the washout of the asymmetry generated by $ \chi^0_1$ and the corresponding term has been considered in the last Boltzman equation as shown in eq. (5) later.

At low energy, some $L$ conserving scatterings could  result in damping of asymmetry.  In $R$-violating MSSM such $L$ conserving processes could be $\chi_i^0 \chi_i^0 \rightarrow W^+ W^-$ through chargino as mediator  and $\chi_i^0 \chi_j^0 \rightarrow l \bar{l}(\nu \bar{\nu})(q \bar{q})$ through left and right charged slepton (sneutrino) (squark) as mediator. Both types of  scattering processes are `self-quenching' \cite{early} and  Boltzman-suppressed with respect to decay process generating asymmetry at low temperature where  out of equilibrium condition is satisfied. This can be seen in the Boltzman equations where thermally averaged  scattering cross-sections are multiplied by the neutralino number densities also. However, without gaugino mass condition, the first
process still could significantly damp the asymmetry by reducing the number density of next to lightest neutralino. This is because
it would be mediated by chargino which could be not so heavy (around TeV range like neutralino) in contrast to the second  process as well as the 
asymmetry generating decay process which are mediated by heavy slepton. With the gaugino mass condition as stated earlier after eq.(2), the first process would not be possible for next to lightest neutralino $\chi_1^0$ which is creating leptonic asymmetry through its decay.  

There are some $L$ violating scatterings which are of two categories :  - (1)  $\chi^0_1l\rightarrow \overline{u} d $  mediated by  left charged slepton with
$\Delta L = -1$, (2) $\chi^0_1\overline{u}\rightarrow l \overline{d} $ 
mediated by left $u$-squark and $ \chi^0_1 d \rightarrow l u  $ mediated by right $d$-squark with $\Delta L = 1$ and their conjugate processes.  The couplings involved in these  for $\chi^0_1$ are constrained by out of equilibrium condition as those are present in $\chi^0_1$ decay. They damp the asymmetry by reducing the number density of $\chi^0_1$.  But they could enhance
or reduce the asymmetry also depending on the sign of $\Delta L$ for such processes. All these scattering processes and their conjugate and inverse processes have been taken into account in the numerical analysis. As $\chi^0_2$ is of lighter mass than $\chi^0_1$ in our case, we have also considered similar scatterings with replacement of $\chi^0_1$  by $\chi^0_2$.

If one considers Feynman diagrams with more than one non-zero $L$ violating couplings  for the neutralino decay there is scope of lightest neutralino 
$ \chi^0_2$ producing
small  $CP$ asymmetry. But because of $\chi_2^0 \chi_2^0 \rightarrow W^+ W^-$  which reduces significantly the number density of $\chi_2^0$, it would have insignificant effect on leptonic asymmetry 
 at  lower scale than that at which $ \chi^0_1$ produces leptonic asymmetry with 
one $L$ violating coupling in the Feynman diagram in figure  \ref{fig1}. However, considering
different one $L$ violating couplings in figure \ref{fig1} one may
consider different decay processes of $ \chi^0_1$ with different charged lepton and quark as decay products to get asymmetry but for simplicity we have assumed except one $L$ violating coupling ($\lambda^{\prime}_{ijk} \equiv \lambda^{\prime}$) the others
are  smaller and ignored in asymmetry evaluation.

 Considering the interference of  tree and next higher order diagram in Fig. \ref{fig1}  one obtains the $CP$ asymmetry parameter $\epsilon$ given by
\begin{equation}
\epsilon = \frac{\Gamma_{\chi^0_1\rightarrow u l\overline{d}}-\Gamma_{\chi^0_1\rightarrow \overline{u} d\overline{l}}}{\Gamma_{\chi^0_1\rightarrow l u\overline{d}}+\Gamma_{\chi^0_1\rightarrow \overline{l} d\overline{u}}+\Gamma_{\chi^0_1\rightarrow l\overline{l}\chi^0_2}+\Gamma_{\chi^0_1\rightarrow q\overline{q}\chi^0_2}}, 
\label{asym}
\end{equation}
where the denominator corresponds to total decay width of next to lightest neutralino. The numerator depends on $CP$ violating phase $\phi $ and one $L$ violating coupling   as 
$$ \left( \Gamma_{\chi^0_1\rightarrow u l\overline{d}}-\Gamma_{\chi^0_1\rightarrow \overline{u} d\overline{l}} \right)
 \propto  |\lambda^{\prime}|^2\sin 2\phi . 
$$
The exact expressions of the numerator and denominator of $\epsilon$ are given in Appendix \ref{sec:appendix} in terms of the integrals which includes  the factor due to  three body phase space.  In the numerator 
interference term from figure \ref{fig1} has been taken into account in which difference of two interference terms for the decay process and its conjugate process contain the imaginary part of the loop integrals associated with higher order diagram in Figure 1. 

We have considered the thermal masses of leptons and quarks above electroweak scale given by \cite{thermal} 
\begin{equation}
m_l(z) = m_\nu(z) = \sqrt{\frac{3}{23}g^2+\frac{1}{32}g'^2}\; \frac{m_{\chi_1}}{z} \; ; \nonumber
\end{equation}
\begin{equation}
m_u(z) = m_d(z)= \sqrt{\frac{1}{3}g_s+\frac{3}{16}g^2+\frac{1}{144}g'^2} \; \frac{m_{\chi_1}}{z}. \quad  
\end{equation}
where $z = \frac{m_{\chi_1}}{T}$, $g = \frac{e}{\sin \theta_W}$, $g' = \frac{e}{\cos \theta_W}$ and
$g_s$ is the strong coupling constant. 

\section{Estimating leptonic asymmetry and neutrino mass}
$\chi^0_{3,4}$ will have insignificant role in
the generation of asymmetry  and $\chi^0_1$ will play the major role as discussed earlier. However, the evolution of the number density of $\chi^0_{ 1}$
will depend on that of $\chi^0_{2}$ through co-annihilation channel. The 
ratio of number densities of $\chi_1$, $\chi_2$ with respect to entropy density and the lepton asymmetry with respect to entropy density are defined as
\begin{align*}
Y_{\chi_i} = \frac{n_{\chi^0_i}(z)}{s(z)} \; ; \; \;
Y_{\Delta L} = \frac{n_{l}(z)-n_{\overline{l}}(z)}{s(z)} 
\end{align*}
respectively    where the entropy density $s(z) = g_*\frac{2\pi^2}{45}\frac{m^3_{\chi_1}}{z^3}$ with 
the effective number of degrees of freedom $ g_* \sim 228$ in $R$ violating MSSM. The coupled Boltzmann equations for $Y_{\chi^0_1}$,$Y_{\chi^0_2}$ and $Y_{\Delta L}$ are :
\begin{widetext}
\begin{align}
\frac{dY_{\chi^0_1}(z)}{dz} &= -\frac{1}{s(z)H(z)z}\bigg[\bigg(\frac{Y_{\chi^0_1}(z)}{Y^{eq}_{\chi^0_1}}-1\bigg)\bigg(\gamma^D_{\chi^0_1} + 2(\gamma_{\chi^0_1l\rightarrow d \overline{u} } + \gamma_{\chi^0_1\overline{u}\rightarrow l \overline{d}} +\gamma_{\chi^0_1 d \rightarrow l u})\bigg) + \sum_{i=1}^2\bigg(\frac{Y_{\chi^0_1}(z)}{Y^{eq}_{\chi^0_1}}\frac{Y_{\chi^0_i}(z)}{Y^{eq}_{\chi^0_i}}-1\bigg)\gamma_{\chi^0_1\chi^0_i\rightarrow f\overline{f}}\bigg] \; ;\nonumber \\
\frac{dY_{\chi^0_2}(z)}{dz} &= -\frac{1}{s(z)H(z)z}\bigg[\bigg(\frac{Y_{\chi^0_2}(z)}{Y^{eq}_{\chi^0_2}}-1\bigg)\bigg(\gamma^D_{\chi^0_2} + 2(\gamma_{\chi^0_2l\rightarrow \overline{u} d} + \gamma_{\chi^0_2\overline{u}\rightarrow l \overline{d}} +\gamma_{\chi^0_2 d \rightarrow l u})\bigg) + \sum_{i=1}^2\bigg(\frac{Y_{\chi^0_2}(z)}{Y^{eq}_{\chi^0_2}}\frac{Y_{\chi^0_i}(z)}{Y^{eq}_{\chi^0_i}}-1\bigg)\gamma_{\chi^0_2\chi^0_i\rightarrow f\overline{f}} \; \nonumber \\
&+ \bigg(\frac{Y^2_{\chi^0_i}(z)}{Y^{eq2}_{\chi^0_2}}-1\bigg)\gamma_{\chi^0_2\chi^0_2\rightarrow W^+W^-} \bigg] \; ;\nonumber \\
\frac{dY_{\Delta L}(z)}{dz} &= \frac{1}{s(z)H(z)z}\left[\epsilon_i\gamma^D_{\chi^0_1}\bigg(\frac{Y_{\chi^0_1}(z)}{Y^{eq}_{\chi^0_1}}-1\bigg) - \sum_{i=1}^2\frac{Y_{\Delta L}(z)}{Y^{eq}_l}\left(\frac{1}{2}\gamma^{LV}_{\chi^0_i} +  \frac{Y_{\chi^0_i}(z)}{Y^{eq}_{\chi^0_i}}\gamma_{\chi^0_i l\rightarrow \overline{u} d} + \gamma_{\chi^0_i\overline{u}\rightarrow l \overline{d}} +\gamma_{\chi^0_i d \rightarrow l u}  \right)\right] \;  \label{eq:boltz}
\end{align}
\end{widetext}
in which   Hubble rate $H(z)=
 \sqrt{\frac{4\pi^3g_*}{45}}\frac{m^2_{\chi_1}}{m_{pl}z^2}$
with planck mass
$m_{pl} = 1.22 \times 10^{19}$ GeV and other quantities are 
\begin{widetext}
\begin{align}
\gamma^D_{\chi^0_1} &= n^{eq}_{\chi^0_1}\frac{K_1(z)}{K_2(z)}\bigg(\Gamma_{\chi^0_1\rightarrow l u\overline{d}}+\Gamma_{\chi^0_1\rightarrow \overline{l} d\overline{u}}+\Gamma_{\chi^0_1\rightarrow l\overline{l}\chi^0_2}+\Gamma_{\chi^0_1\rightarrow q\overline{q}\chi^0_2}\bigg)   ; \;
\gamma^D_{\chi^0_2} = n^{eq}_{\chi^0_2}\frac{K_1(z)}{K_2(z)}\bigg(\Gamma_{\chi^0_2\rightarrow l u\overline{d}}+\Gamma_{\chi^0_2\rightarrow \overline{l} d\overline{u}}\bigg) =\gamma^{LV}_{\chi^0_2}  \; ;  \nonumber\\
\gamma^{LV}_{\chi^0_1} &= n^{eq}_{\chi^0_1}\frac{K_1(z)}{K_2(z)}\bigg(\Gamma_{\chi^0_1\rightarrow l u\overline{d}} +\Gamma_{\chi^0_1\rightarrow \overline{l} d\overline{u}}\bigg)   ;\;
\gamma_{\psi_1\psi_2\rightarrow \phi_1\phi_2} = \frac{m_{\chi^0_2}}{64\pi^4 z}\int^{\infty}_{s_{min}}ds \frac{2\lambda(s,m^2_{\psi_1},m^2_{\psi_2})}{s}\sigma(s)\sqrt{s}K_1\bigg(\frac{\sqrt{s}z}{m_{\chi^0_2}}\bigg)  ;\;\nonumber \\
s_{min} &= max[(m_{\psi_1}+m_{\psi_2})^2,(m_{\phi_1}+m_{\phi_2})^2] \nonumber.
\end{align}
\end{widetext}
In first two Boltzman equations $f$ corresponds to leptons and quarks.  $K_i(z)$ are usual modified Bessel Functions and different  thermally averaged decays are defined as
$\gamma^D_{\chi^0_2},\gamma^D_{\chi^0_1} $, and $L$ violating   $\gamma^{LV}_{\chi^0_1} \; \textrm{and} \; 
\gamma^{LV}_{\chi^0_2}$ and different thermally averaged scattering cross-sections are defined as $
\gamma_{\psi_i\psi_j\rightarrow \phi_1\phi_2}$ with $i,j=1,2$ and  $\epsilon$ is the $CP$ asymmetry parameter as mentioned in eq.\eqref{asym}. The superscript $eq$ denotes the corresponding values in thermal equilibrium. Unlike $\gamma^D_{\chi^0_2}$, the
$\gamma^D_{\chi^0_1}$ has $L$ conserving decay term also and unlike $\chi^0_1$, the $\chi^0_2$ has annihilation to $W^+ W^-$ also.  Both $L$ conserving and $L$ violating scatterings
are considered in first two Boltzman equations as they reduce $n_{\chi^0_i}$  while  $L$ violating scatterings  are considered in the 
last Boltzman equation.   The thermally averaged $\gamma$ corresponding to various scatterings processes, their conjugate
and inverse processes are same and not written separately but associated factors with $\gamma$ have been taken appropriately in Boltzman equations.

In presence of sphalerons the leptonic asymmetry $Y_{\Delta L}$ will be converted to baryonic asymmetry $Y_{\Delta B}$ as \cite{saphos}
\begin{align*}
Y_{\Delta B} = \frac{n_{B}(z)-n_{\overline{B}}(z)}{s(z)} =  - \left( \frac{8 N_f + 4 N_H}{ 22 N_f + 13 N_H} \right) Y_{\Delta L}
\end{align*}
where $N_f =3$ is the number of lepton  generations and $N_H =2$ is the number of Higgs doublets in the $R$-parity violating MSSM.

\begin{widetext}
\begin{table}[!ht]
    \centering
    \begin{tabular}{|c|c|c|c|c|c|c|c|c|c|c|} \hline \hline 
    & BP1 & BP2 & BP3 & BP4 & BP5 & BP6 & BP7 & BP8 & BP9 & BP10 \\ \hline
      $m_{\chi_1}$   & 5 TeV & 5 TeV & 5 TeV & 5 TeV & 5 TeV & 6 TeV & 6 TeV & 6 TeV & 3 TeV & 3 TeV \\
      $m_{\chi_2}$   & 4 TeV & 4 TeV & 4 TeV & 4 TeV & 3.5 TeV & 5 TeV & 5 TeV & 4 TeV & 2 TeV & 1.5 TeV \\
      $\lambda'$  & 0.01 & 0.01 & 0.01 & 0.1 & 0.01 & 0.01 & 0.01 & 0.01 & 0.01 & 0.01 \\
      $m_{\tilde{f}}$   & 60 TeV & 30 TeV & 7 TeV & 6 TeV & 60 TeV & 7 TeV & 30 TeV & 7 TeV & 5 TeV & 4 TeV \\ \hline
    \end{tabular}
    \caption{Different benchmark points considered for different plots in Figure \ref{fig3} (a) , (b) and (c). For all these sets $\phi = \pi/4$.}
    \label{tab:BP}
\end{table}   
\end{widetext}

\begin{widetext}
\begin{table}[!ht]
    \centering
    \begin{tabular}{|c|c|c|c|c|c|c|c|c|} \hline \hline 
    & BP11 & BP12 & BP13 & BP14 & BP15 & BP16 & BP17 & BP18 \\ \hline
      $m_{\chi_1}$   & 6 TeV & 6 TeV & 5 TeV & 5 TeV & 6 TeV & 5 TeV & 5 TeV & 5 TeV \\
      $m_{\chi_2}$   & 5 TeV & 5 TeV & 4 TeV & 4 TeV & 5 TeV & 4 TeV & 4 TeV & 4 TeV\\
      $\lambda'$  & 0.01 & 0.01 & 0.01 & 0.01 & 0.01 & 0.1 & 0.1 & 0.1 \\
      $m_{\tilde{q}}$   & 7 TeV & 30TeV & 30 TeV  & 6 TeV & 20 TeV  & 30 TeV & 6 TeV  & 30 TeV \\ 
      $m_{\tilde{l}}$   & 7 TeV & 30 TeV & 6 TeV  & 30 TeV & 7 TeV & 6 TeV  & 30 TeV & 6 TeV \\
      $\phi$ &  $ \pi/8$  &  $\pi/8$  &  $\pi/4$  &   $\pi/4$ &  $\pi/8$ & $ \pi/8 $ & $\pi/8$ & $ \pi/6$\\ \hline
    \end{tabular}
    \caption{Different benchmark points considered for different plots in Figure \ref{fig3} (d) and (e).}
    \label{tab:BP1}
\end{table} 
\end{widetext}

\begin{widetext}
\begin{figure}[th] 
\centering
\begin{tabular}{ccc}
\epsfig{file=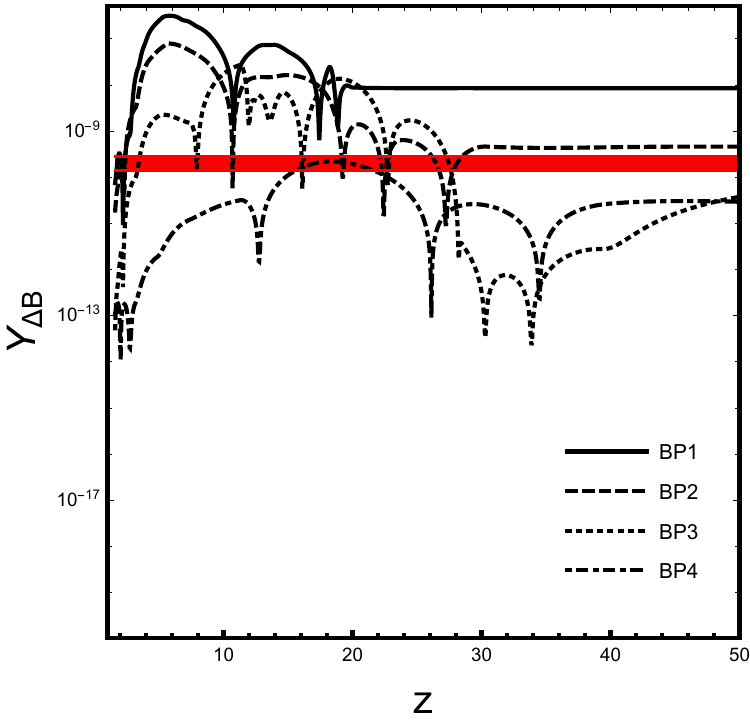,width=0.66\linewidth,clip=}&
\epsfig{file=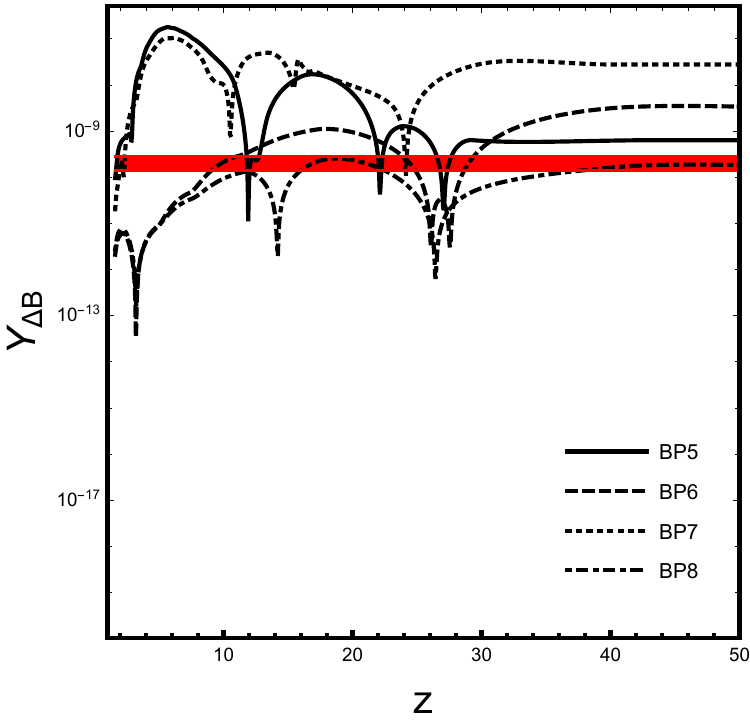,width=0.66\linewidth,clip=}&
\epsfig{file=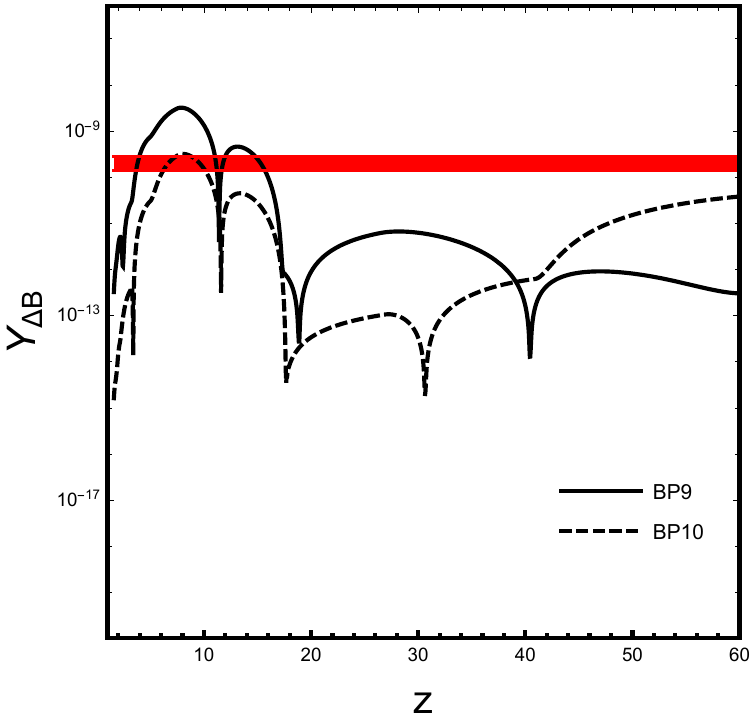,width=0.66\linewidth,clip=}\\
(a)&(b)&(c)
\end{tabular}
\begin{tabular}{cc}
\epsfig{file=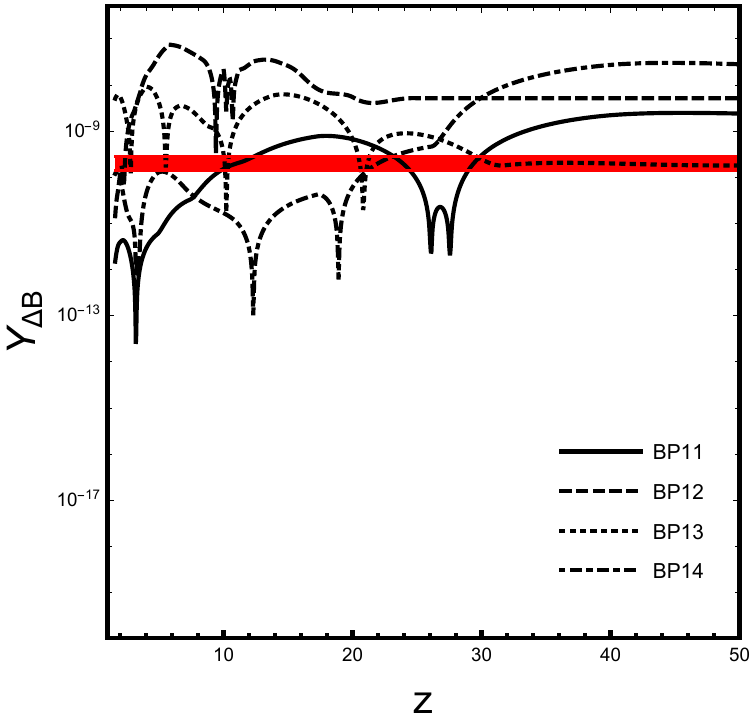,width=0.66\linewidth,clip=}&
\epsfig{file=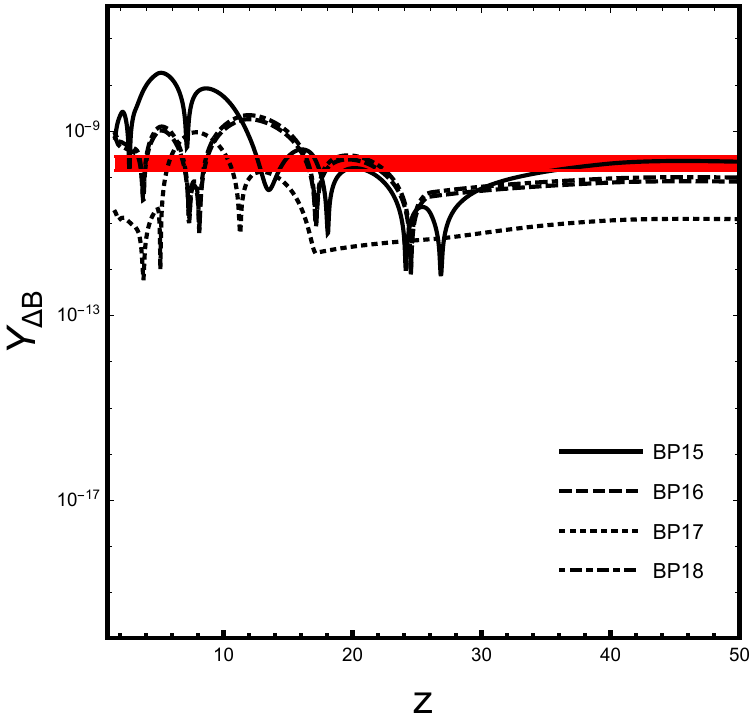,width=0.66\linewidth,clip=}\\
(d)&(e)
\end{tabular}
\caption{ Plot of $Y_{\Delta B}$ versus $z=m_\chi/T$ shown in  (a), (b) and (c) with  phase $\phi = \pi/4$ for different sets of   benchmark points as shown in Table \ref{tab:BP}. Plot shown in (d) and (e) for different sets of   benchmark points as shown in Table \ref{tab:BP1}.
The red horizontal line  corresponds to required approximate baryonic asymmetry. }
\label{fig3}
\end{figure}
\end{widetext}

Based on recent experimental data \cite{asym} the observed baryonic asymmetry corresponds to $Y_{\Delta B} \approx  10^{-11}$ at recombination. In Fig. \ref{fig3}, we have shown the evolution of the baryonic asymmetry $Y_{\Delta B}$ with temperature ($z = m_{\chi_1}/T$) after numerically solving above  Boltzmann equations. The required  $Y_{\Delta B}$ could be  obtained around $z \sim 30$ at leptogenesis scale and to get
$Y_{\Delta B}$ at recombination, the $Y_{\Delta B}$ at leptogenesis scale shown in figure \ref{fig3} is to be divided by the additional entropy dilution factor $f \approx 30 $ \cite{early}.    For numerical analysis, all left and right squark and all left and right slepton masses have been assumed to be equal. We have considered different sets of benchmark points in Table I and Table II. In Table I, we have considered slepton and squark masses to be equal and such equal mass has been denoted as $m_{\tilde{f}}$ . In Table I, for all sets, the phase $\phi = \pi/4$ has been considered for which lepton asymmetry is maximal as it is proportional to $\sin{2 \phi}$. However, in Table II, we have considered different slepton and squark masses and also different phases in different sets of benchmark points. Figure 3 (a), (b) and (c) correspond to benchmark points in Table I and Figure 3 (d) and (e) correspond to benchmark points in Table II. 
 Choice of $\lambda^{\prime}$ in all cases satisfies  phenomenological constraint  \cite{cons} for heavy sfermion masses considered here. $|\mu| $ parameter is required to be higher than the two masses of lighter two neutralinos as discussed earlier. We will consider its specific choice later in our discussion for neutrino mass.  As one can see in the last term of the 3rd Boltzman equations, there are particularly three different scattering processes mentioned whose inverse also has been taken into account with appropriate pre-factors - which could play some role in washing out or increasing the asymmetry. The first one depends on slepton mass and the other two depends on  squark masses. 
 To study the role of such terms in generating asymmetry we have considered same as well as different choices of slepton and squark masses in Table I and II.

 The required minimum freeze out temperature  is $ T_{\rm out} \sim 200$ GeV for 
sphaleron  to convert leptonic asymmetry to baryonic asymmetry  \cite{Rubakov:1996vz,Hambye:2000zs}  for weakly first order or second order phase transition. We do not need here strictly first order phase transition. So the leptonic asymmetry can be successfully converted to baryonic asymmetry by the sphalerons for freeze out temperature $T \gtrsim 200$ GeV for 
which $z \lesssim 30$ depending on our choices of 
$m_{\chi_1}$ in Table I and II. If this condition is not satisfied in any plot in Figure 3, we have to discard that for successful generation of baryonic asymmetry. This is because the conversion of leptonic asymmetry to baryonic asymmetry will be actually very less than that shown in the figure as the relation between these two asymmetries as mentioned earlier through sphaleron transition will not be valid. As for example, in the plot labelled as BP8 in Figure 3 (b), the required baryonic asymmetry seems to be obtained but  freeze-out occurs at $z \sim 40 $ and this means the freeze out temperature is below 200 GeV. So we have to discard set of benchmark points in BP8 for successful generation of baryonic asymmetry. From Figure 3 (a), (b) and (c), it is found that for benchmark points BP1, BP2, BP5, BP6, BP7, the successful baryonic asymmetry could be obtained. However, for BP2 and BP5, the above condition is narrowly satisfied as $z \lesssim 25 $ is expected for freeze out for $m_{\chi_1} = 5$ TeV. So the lightest neutralino mass $m_{\chi_2}$ could be about 5 GeV and sfermion mass about 7 GeV as seen in BP6. If $m_{\chi_2}$ is considered futher lower at about 4 GeV, then the sfermion mass  could be higher than  30 TeV or so as seen in plots BP1 and BP2. In this case, for lower sfermion mass, the required asymmetry is not obtained as can be seen in other plots in Figure 3 (a), (b) and (c). In Figure 3, (d), plots BP11, BP12, BP13 and BP14 satisfy the required freeze out value of $z$ and may be considered for the generation of required baryonic asymmetry. But in Figure 3 (e), all the plots are not suitable for the generation of baryonic asymmetry due to lower than 200 GeV freeze out temperature with corresponding higher $z$ value. Plots BP11 and BP12 have the same benchmark points like BP6 and BP7 respectively except the phase which is $\pi/8$. There is no significant change due to that. However, plots in BP13 and BP14 are somewhat interesting. In BP13, the squark mass is heavier and in that case, for lighter slepton mass at 6 TeV, the required baryonic asymmetry is obtained. In BP14, the slepton mass is heavier and in that case, for lighter squark  mass at 6 TeV, the required baryonic asymmetry is obtained. In both BP13 and BP14, the lightest neutralino mass is at 4 GeV. In Figure 3 (e), particularly in case of BP15, with light slepton but squark mass at 20 TeV, required asymmetry is not produced. From earllier plots in Figure 3 (d), it seems further higher squark mass like 30 TeV or further lower like 7 TeV could work. In case of BP16, BP17 and BP18, we have considered higher value of $L$ violating coupling, in comparison to other sets of benchmark points and required baryonic asymmetry is not obtained. So from Figure 3,  it is seen that for successful leptogenesis at low energy scale around TeV, the allowed parameter space of sfermion masses and neutralino masses get constrained from the requirement of conversion of leptonic asymmetry to baryonic asymmetry through sphaleron transition. From different plots, one can see that with appropriate combination of other parameters, the successful baryonic asymmetry could be possible for (1) lightest neutralino mass as low as 4 TeV (2)  slepton or squark mass as low as  6 TeV for which either squark or slepton is required to be heavier around 30 TeV (3) equal squark and slepton mass  as low as 7 TeV.
 
The same $L$ violating $\lambda^{\prime}$ coupling considered for leptogenesis could also give Majorana neutrino mass. It can be obtained from one loop Feynman diagram with left and right $d$-squark and  $d$-quark (with left-right mixing) in the loop with $L$ violating coupling. Instead of see-saw mechanism here smallness of neutrino mass occurs due to one loop diagram. Neutrino mass matrix elements 
can be written as \cite{gordana}
\begin{equation}
m_{\nu_{ij}} \approx \sum_{k,l} \frac{3 \;\lambda^{\prime}_{ikl}\lambda^{\prime}_{jlk}m_km_l \widetilde{m}}{16\pi^2\widetilde{m}_q^2}
\end{equation}
where $m_k$ and $m_l$ are the masses of $d$-quarks and $\widetilde{m}  \sim (A, \mu)$ corresponds to SUSY breaking parameters and
$ \widetilde{m}_q$ is the $d$-squark masses. Main contribution is expected to come from $b$-quark mass in the numerator.    Considering one of the $\lambda^{\prime} \sim  10^{-2}$ corresponding to  Figure \ref{fig3} (for which successful baryonic asymmetry has been obtained) and other $\lambda^{\prime}$ couplings  lesser than that and $\widetilde{m} = 20 \;{\rm TeV}; \quad \widetilde{m}_q = 30 \;{\rm TeV}$ and  $k = l = b$ one can easily satisfy the cosmological bound $\sum m_{\nu_i} \lesssim 0.1 $ eV \cite{cosmo}. Considering $\lambda^{\prime} \sim  10^{-2}$ is  appropriate to get mass square differences of about $2 \times 10^{-3} {\mbox{eV}}^2$ to satisfy atmospheric neutrino oscillation data  \cite{olive}.  For other $L$ violating couplings like $\lambda^{\prime}_{1jk}, \lambda^{\prime}_{2jk}$  lesser than $\lambda^{\prime}$  it is possible to get mass square differences of about $7 \times 10^{-5}$  satisfying 
solar  neutrino oscillation data \cite{olive}. If we consider lighter squark mass about 6 TeV as follows from   Figure 3 (d) , then to get similar order of neutrino mass with one $\lambda^{\prime}$ value as $10^{-2}$ like before and other $\lambda^{\prime}$ couplings lesser than that, we may consider either $k$ or $l$ to correspond to $d$ quark instead of $b$ quark, then it could be possible to satisfy required neutrino masses square differences and the cosmological bound on neutrino masses. However, alternate option of reducing 
 $\widetilde{m}$ to about 1 TeV is not possible in our mechanism of generating baryonic asymmetry because that would violate our requirement of gaugino mass condition mentioned after eq. (2). 
 
\section{Conclusion} The lighter two neutralinos are mostly gauginos   after symmetry breaking at the electro-weak scale. This could be envisioned in Supergravity models with symmetry broken in the hidden sector \cite{msugra}. At LHC there is scope to verify the possibility of such leptogenesis mechanism for which there could be pair production of such gauginos followed by decay of next to lightest neutralino to the lightest one by  $L$ conserving decay mode and the lightest and next to lightest one by
$L$ violating decay mode as mentioned earlier giving multi-lepton signature. Particularly, the case of either slepton or squark mass  as light as 6 TeV is possible  in the plots in Figure 3 (d).  In that case, their pair production may be possible at LHC in near future. However, in several sets of benchmark points, the sfermion masses are found to be heavier.
Such heavy masses are consistent \cite{ibanez} with the kind of Higgs mass observed at LHC.   Besides, because of  sfermion masses in few TeV scale or more, it is
difficult to constrain  Majorana phases \cite{phase} even after recent improvement on the experimental bound 
of electric dipole moment of particularly electron \cite{edme}. So the phase $\phi = \pi/4$ 
as considered in our numerical analysis, is easily allowed.  However, following various cases considered in Figure 3  it seems that the lightest supersymmetric particle - the lightest neutralino in MSSM scenario should have mass not lower than about 4 TeV. 

In this scenario in $\slashed{R}$ MSSM  with gaugino mass conditions, baryonic asymmetry  could be  sufficient without any fine tuning of model parameters because leptogenesis   is possible with one $L$ violating coupling and  $CP$ violation due to Majorana phase could be maximal.  

\begin{center}
 {\bf Acknowledgments}   
\end{center}
We like to thank the organisers of WHEPP-XIV held at IIT, Kanpur for inviting us where  part of the work was completed. We are  grateful to the referee for various valuable suggestions for improving the work.

\hspace*{\fill}

\appendix

\section{Details of Asymmetry}
\label{sec:appendix}

Here we present the detailed expression for the asymmetry given in eq.\eqref{asym} related to leptonic asymmetry. The asymmetry $\epsilon$ can be rewritten as:
\begin{align}
   \epsilon &= \frac{\Gamma_{\chi^0_1\rightarrow u l\overline{d}}-\Gamma_{\chi^0_1\rightarrow \overline{u} d\overline{l}}}{\Gamma_{\chi^0_1\rightarrow l u\overline{d}}+\Gamma_{\chi^0_1\rightarrow \overline{l} d\overline{u}}+\Gamma_{\chi^0_1\rightarrow l\overline{l}\chi^0_2}+\Gamma_{\chi^0_1\rightarrow q\overline{q}\chi^0_2}} = \frac{\Gamma_\delta}{\Gamma_{tot}}
\end{align}
where the numerator is given as:
\begin{widetext}
\begin{align}
\Gamma_\delta &= \Gamma_{\chi^0_1\rightarrow u l\overline{d}}-\Gamma_{\chi^0_1\rightarrow \overline{u} d\overline{l}} = \int_{z_{l_{-}}}^{z_{l_{+}}} \int_{z_{u_{-}}}^{z_{u_{+}}} \frac{3m_{\chi_1}}{256\pi^3}\left(\frac{1}{2}\sum_{\rm spins}|\delta|^2 \right) dz_l dz_d \\
   \mbox{in which } \; z_{l_{-}} &= 2r_l ;\quad z_{l_{+}} = 1 + r^2_l - (r_u + r_{\bar{d}})^2; \quad r_i = m_i/m_{\chi^0_1}; \quad z_f = 2p_{\chi^0_1}.p_f/(m^2_{\chi^0_1})= 2E_f/m_{\chi^0_1} \nonumber \\
    \mbox{and}\; (z_{u})_\pm &= \frac{1}{2(1-z_l+r^2_l)}\left[(2-z_l)(1+r^2_l + r^2_u - r^2_{\bar{d}} - z_l) \pm \sqrt{z^2_l - 4r^2_l}\Lambda^{1/2}(1+r^2_l-z_l,r^2_u,r^2_{\bar{d}})\right] \nonumber \\
     \mbox{in which} \; \Lambda(x,y,z) &= x^2 + y^2 + z^2 - 2xy - 2xz - 2yz \quad  z_{\bar{d}} = 2 - z_u - z_l \; . \nonumber \\
   \mbox{ $|\delta|^2$ in  $\Gamma_\delta $ is given by}  \nonumber \\
  |\delta|^2 &= -4r_{\chi^0_2}\; \lambda^{\prime 2}_{lud}\; \sin 2\phi\; g^2g'^2\left[\Im[\tilde{C}^l_1]\left(\frac{1}{4}\frac{z_l(1+r^2_l - r^2_u - r^2_{\bar{d}} - z_l)}{(1+r^2_l-z_l - r^2_{\tilde{l}})^2} \right. \right. \nonumber \\ &+ \frac{1}{6}\frac{r^2_l(1 + r^2_l-r^2_u-r^2_{\bar{d}} - z_l)}{(1+r^2_u -r^2_{\tilde{u}}-z_u)(1+r^2_l-r^2_{\tilde{l}}-z_l)}   \nonumber \\
    &+ \frac{1}{6}\frac{z_u(1+r^2_u-r^2_l-r^2_{\bar{d}}-z_u) - z_{\bar{d}}(1+r^2_{\bar{d}}-r^2_u-r^2_l-z_{\bar{d}}) + z_l(1+r^2_l - r^2_u - r^2_{\bar{d}} - z_l)}{(1+r^2_u -r^2_{\tilde{u}}-z_u)(1+r^2_l-r^2_{\tilde{l}}-z_l)} \nonumber \\
    &+ \left.\frac{1}{3}\frac{z_{\bar{d}}(1+r^2_{\bar{d}} - r^2_u -r^2_l-z_{\bar{d}} - z_u(1+r^2_u-r^2_l-r^2_{\bar{d}}-z_u) + z_l(1+r^2_l-r^2_u-r^2_{\bar{d}}-z_l)}{(1+r^2_u -r^2_{\tilde{u}}-z_u)(1+r^2_{\bar{d}}-r^2_{\tilde{d}}-z_{\bar{d}})}\right)\nonumber \\
    &+\Im[\tilde{C}^l_2]\left(\frac{1}{2}\frac{r^2_l(1+r^2_l+r^2_u-r^2_{\bar{d}}-z_l)}{(1+r^2_l-z_l-r^2_{\tilde{l}})^2} + \frac{1}{3}\frac{r^2_l(1+r^2_l-r^2_u-r^2_{\bar{d}}-z_l)}{(1+r^2_l-z_l-r^2_{\tilde{l}})(1+r^2_{\bar{d}}-r^2_{\tilde{d}}-z_{\bar{d}})}\right) \nonumber \\
    &+ \Im[\tilde{C}^u_1]\left(\frac{1}{18}\frac{z_u(1+r^2_u-r^2_l-r^2_{\bar{d}}-z_u)-z_{\bar{d}}(1+r^2_{\bar{d}}-r^2_u-r^2_l-z_{\bar{d}})+z_l(1+r^2_l-r^2_u-r^2_{\bar{d}}-z_l)}{(1+r^2_l-z_l-r^2_{\tilde{l}})(1+r^2_u-z_u-r^2_{\tilde{u}})}\right. \nonumber \\
    &-\frac{1}{108}\frac{z_u(1+r^2_u-r^2_l-r^2_{\bar{d}}-z_u)}{(1+r^2_l-r^2_{\tilde{l}}-z_l)(1+r^2_{\bar{d}}-r^2_{\tilde{r}}-z_{\bar{d}})}\nonumber 
    \end{align}
    \end{widetext}
    \begin{widetext}
    \begin{align}
     &+ \left.\frac{1}{27}\frac{z_{\bar{d}}(1+r^2_{\bar{d}}-r^2_u-r^2_l - z_{\bar{d}}) - z_l(1+r^2_l-r^2_u-r^2_{\bar{d}}-z_l) + z_u(1+r^2_u-r^2_l-r^2_{\bar{d}}-z_u))}{(1+r^2_u-r^2_{\tilde{u}}-z_u)(1+r^2_{\bar{d}}-r^2_{\tilde{d}}-z_{\bar{d}})}\right) \nonumber\\
    &+\left.\Im[\tilde{C}^u_2]\left(\frac{1}{18}\frac{r^2_u(1+r^2_u-r^2_l-r^2_{\bar{d}}-z_u)}{(1+r^2_l-r^2_{\tilde{u}}-z_l)(1+r^2_u-r^2_{\tilde{u}}-z_u)}+\frac{1}{27}\frac{r^2_u(1+r^2_u-r^2_l-r^2_{\bar{d}}-z_u)}{(1+r^2_u-r^2_{\tilde{u}}-z_u)(1+r^2_{\bar{d}}-r^2_{\tilde{d}}-z_{\bar{d}})}\right)\right] 
    \end{align}
    \end{widetext}
    in which
    \begin{widetext}
    \begin{align}
    \tilde{C}^f_1 &= \frac{1}{(4r^2_f-2z_f)}\left[2r^2_f\left(\mathcal{B}_0(m^2_f,m^2_{\tilde{f}},m^2_{\chi^0_2})-\mathcal{B}_0(m^2_{\chi^0_1}+m^2_f-m^2_{\chi^0_1}z_f,m^2_{\chi^0_2},m^2_f) \right. \right. \nonumber \\
    &+ \left.(m^2_f-m^2_{\tilde{f}})\mathcal{C}_0(m^2_{\chi^0_1},m^2_f,\frac{1}{2}m^2_{\chi^0_1}z_f,m^2_{\tilde{f}},m^2_f,m^2_{\chi^0_2})\right) \nonumber \\
    &- z_f\left(\mathcal{B}_0(m^2_{\chi^0_1},m^2_{\tilde{f}},m^2_f)- \mathcal{B}_0(m^2_{\chi^0_1}+m^2_f-m^2_{\chi^0_1}z_f,m^2_{\chi^0_2},m^2_{\tilde{f}})\right.  \nonumber \\
    &+ \left.\left.(m^2_{\chi^0_2}-m^2_{\tilde{f}})\mathcal{C}_0(m^2_{\chi^0_1},m^2_f,\frac{1}{2}m^2_{\chi^0_1}z_f,m^2_{\tilde{f}},m^2_f,m^2_{\chi^0_2})\right)\right]; \nonumber \\ 
        \tilde{C}^f_2 &= \frac{1}{(4r^2_f-2z_f)}\left[2r^2_f\left(\mathcal{B}_0(m^2_{\chi^0_1},m^2_{\tilde{f}},m^2_f)- \mathcal{B}_0(m^2_{\chi^0_1}+m^2_f-m^2_{\chi^0_1}z_f,m^2_{\chi^0_2},m^2_{\tilde{f}}) \right. \right. \nonumber \\
    &+ \left.(m^2_{\chi^0_2}-m^2_{\tilde{f}})\mathcal{C}_0(m^2_{\chi^0_1},m^2_f,\frac{1}{2}m^2_{\chi^0_1}z_f,m^2_{\tilde{f}},m^2_f,m^2_{\chi^0_2})\right) \nonumber \\
    &- z_f\left(\mathcal{B}_0(m^2_f,m^2_{\tilde{f}},m^2_{\chi^0_2})-\mathcal{B}_0(m^2_{\chi^0_1}+m^2_f-m^2_{\chi^0_1}z_f,m^2_{\chi^0_2},m^2_f) \right.  \nonumber \\
    &+ \left. (m^2_f-m^2_{\tilde{f}})\mathcal{C}_0(m^2_{\chi^0_1},m^2_f,\frac{1}{2}m^2_{\chi^0_1}z_f,m^2_{\tilde{f}},m^2_f,m^2_{\chi^0_2})\right)\left.\right] \; ; \nonumber\\  
    \mathcal{B}_0(p^2,m^2_1,m^2_2) &= \int \frac{dl^4}{(2\pi)^4}\frac{1}{(l^2-m^2_1)((l+p)^2-m^2_2)} \; ; \nonumber \\ \mathcal{C}_0(p^2_1,p^2_2,p_1.p_2,m^2_1,m^2_2,m^2_3) &= \int \frac{dl^4}{(2\pi)^4}\frac{1}{(l^2-m^2_1)((l+p_1)^2-m^2_2)((l+p_2)^2-m^2_3)} \; .\nonumber
\end{align}
\end{widetext}
The denominator of Eq. (3) is given as:
\begin{align}
    \Gamma_{tot} &= \Gamma_{\chi^0_1\rightarrow l u\overline{d}}+\Gamma_{\chi^0_1\rightarrow \overline{l} d\overline{u}}+\Gamma_{\chi^0_1\rightarrow l\overline{l}\chi^0_2}+\Gamma_{\chi^0_1\rightarrow q\overline{q}\chi^0_2}
\end{align}
in which 
\begin{widetext}
\begin{align}
    \Gamma_{\chi^0_1\rightarrow l u\overline{d}}+\Gamma_{\chi^0_1\rightarrow \overline{l} d\overline{u}} &= \int_{z_{l_{-}}}^{z_{l_{+}}} \int_{z_{u_{-}}}^{z_{u_{+}}} \frac{6m_{\chi_1}}{256\pi^3}\left(\frac{1}{2}\sum_{\rm spins}|\mathcal{M}_+|^2 \right) dz_l dz_d \\
   \mbox{in which } \; z_{l_{-}} &= 2r_l ;\quad z_{l_{+}} = 1 + r^2_l - (r_u + r_{\bar{d}})^2; \quad z_f = 2p_{\chi^0_1}.p_f/(m^2_{\chi^0_1})= 2E_f/m_{\chi^0_1}\nonumber \\
    \mbox{and}\; (z_{u})_\pm &= \frac{1}{2(1-z_l+r^2_l)}\left[(2-z_l)(1+r^2_l + r^2_u - r^2_{\bar{d}} - z_l) \pm \sqrt{z^2_l - 4r^2_l}\Lambda^{1/2}(1+r^2_l-z_l,r^2_u,r^2_{\bar{d}})\right]. \nonumber \\
   |\mathcal{M}_+|^2 &= \lambda^{\prime 2}_{uld} g^{\prime 2}\left(z_l(1 + r^2_l - r^2_u - r^2_{\bar{d}} - r^2_{\bar{d}} - z_l)\left[\frac{1}{2(1+r^2_l-r^2_{\tilde{l}}-z_l)^2} + \frac{1}{6(1+r^2_l-r^2_{\tilde{l}}-z_l)(1+r^2_u - r^2_{\tilde{u}} - z_u)} \right. \right. \nonumber \\
   &+ \left. \left. \frac{1}{3(1+r^2_l-r^2_{\tilde{l}}-z_l)(1+r^2_{\bar{d}}-r^2_{\tilde{d}}-z_{\bar{d}})}\right] \right. \nonumber \\
   &+ \left. z_u(1 + r^2_u - r^2_l - r^2_{\bar{d}}-  z_u)\left[\frac{1}{18(1+r^2_u-r^2_{\tilde{u}} - z_u)^2} + \frac{1}{6(1 + r^2_u-r^2_{\tilde{u}} - z_u)(1 + r^2_u - r^2_{\tilde{u}}-z_u)} \right. \right. \nonumber \\
   &+ \left. \left.\frac{1}{3(1 + r^2_l-r^2_{\tilde{l}} - z_l)(1 + r^2_{\bar{d}}-r^2_{\tilde{d}} - z_{\bar{d}})} - \frac{1}{9(1 + r^2_u-r^2_{\tilde{u}} - z_u)(1 + r^2_{\bar{d}}-r^2_{\tilde{d}} - z_{\bar{d}})}\right] \right. \nonumber\\
   &+ \left. z_{\bar{d}}(1+r^2_{\bar{d}} - r^2_u - r^2_l - z_{\bar{d}})\left[\frac{2}{9(1+r^2_{\bar{d}} - r^2_{\tilde{r}} - z_{\bar{d}})^2} - \frac{1}{6(1+r^2_l - r^2_{\tilde{l}} - z_l)(1+r^2_u - r^2_{\tilde{u}} - z_u)} \right. \right. \nonumber \\
   &+ \left.\left. \frac{1}{3(1+r^2_l - r^2_{\tilde{l}} - z_l)(1+r^2_{\bar{d}} - r^2_{\tilde{d}} - z_{\bar{d}})} + \frac{1}{9(1+r^2_u - r^2_{\tilde{u}} - z_u)(1+r^2_{\bar{d}} - r^2_{\tilde{d}} - z_{\bar{d}})}\right]\right) 
   \label{eq:A6}
\end{align}
\end{widetext}
and 
\begin{widetext}
\begin{align}
    \Gamma_{\chi^0_1\rightarrow f \bar{f} \chi^0_2} &= \int_{z_{1_{-}}}^{z_{1_{+}}} \int_{z_{2_{-}}}^{z_{2_{+}}} \frac{m_{\chi_1}}{256\pi^3}\left(\frac{1}{2}\sum_{\rm spins}|\mathcal{M}|^2 \right) dz_1 dz_2  \\
   \mbox{in which } \; z_{1_{-}} &= 2r_f ;\quad z_{1_{+}} = 1 + r^2_f - (r_f + r_{\chi^0_2})^2; \quad z_{1(2)} = 2p_{\chi^0_1}.p_{f(\bar{f})}/(m^2_{\chi^0_1})= 2E_{f(\bar{f})}/m_{\chi^0_1}\nonumber \\
    \mbox{and}\; (z_{2})_\pm &= \frac{1}{2(1-z_1+r^2_f)}\left[(2-z_1)(1+2r^2_f  - r^2_{\chi^0_2} - z_1) \pm \sqrt{z^2_1 - 4r^2_f}\Lambda^{1/2}(1+r^2_f-z_1,r^2_f,r^2_{\chi^0_2})\right] ;\nonumber \\
    |\mathcal{M}|^2 &= N^f_c4g^2g^{\prime 2}T^{f2}_3(Q_f-T^f_3)^2\left[\frac{z_1(1-r^2_{\chi^0_2}-z_1)}{(1+r^2_f-r^2_{\tilde{f}}-z_1)^2} + \frac{z_2(1-r^2_{\chi^0_2}-z_2)}{(1+r^2_f-r^2_{\tilde{f}}-z_2)^2} \frac{r_{\chi^0_2}(r^2_{\chi^0_2}-2r^2_f+z_1+z_2-1)}{(1+r^2_f-r^2_{\tilde{f}}-z_1)(1+r^2_f-r^2_{\tilde{f}}-z_2)}\right]\nonumber 
\end{align}
\end{widetext}
where the factor $N^f_c$ (equal to 1 for leptons and 3 for quarks) is for the sum over colors and $Q_f$ is the electric charge of the fermion (i.e for charged leptons $Q_l = -1$) and $T^f_3=1/2$ for $f = u,\nu$ and $T^f_3=-1/2$ for $f = d,l$. In Figure 1, there are three diagrams at the tree level and two diagrams at the loop level.   For brevity, the decay width of $\chi^0_1$ has been shown for one tree level diagram with $\widetilde{l}$ in the propagator and $\Gamma_\delta$ has been shown for the interference of one tree diagram and one loop diagram with $\widetilde{l}$ in the propagators. For  $\widetilde{u}$ and  $\widetilde{d}$ in the propagators one will get similar expressions with the appropriate replacement of  $\widetilde{l}$ by $\widetilde{u}$ or $\widetilde{d}$ by their masses and appropriate couplings. For equal slepton and squark mass, all these diagrams will give $\Gamma_\delta$ approximately 6 times than what is shown and the decay width of $\chi^0_1$ will be around 9 times than what is shown.   

\end{document}